\documentclass[traditabstract,longauth]{aa}  
\makeatletter
\def\@LN#1#2{} 
\def\@LN@col#1{} 
\let\linenumbers\relax
\let\nolinenumbers\relax
\makeatother

\usepackage[T1]{fontenc}
\usepackage{txfonts}
\usepackage{graphicx}
\usepackage[dvipsnames]{xcolor}
\usepackage{amsmath,amssymb}
\usepackage{lscape}
\usepackage{subcaption}
\usepackage{placeins}
\usepackage{hyperref}
\hypersetup{
    colorlinks=true,
    citecolor={blue}
}

\usepackage{natbib}
\bibpunct{(}{)}{;}{a}{}{,} 

\titlerunning{Thermal emission of old neutron stars with rotochemical heating}

\authorrunning{Rodríguez et al.}

\begin{document} 

\title{Explaining the thermal emission of old neutron stars with rotochemical heating and magnetized superconducting protons}

\author{
Luis E. Rodr\'iguez\thanks{e-mail: lsrodrig@uc.cl}\inst{1}
\and Andreas Reisenegger\inst{2,3}
\and Denis Gonz\'alez-Caniulef\inst{4}
\and Crist\'obal Petrovich\inst{1,5}
}

\institute{
Instituto de Astrof\'isica, Pontificia Universidad Cat\'olica de Chile, Av. Vicu\~na Mackenna 4860, Macul, Santiago, Chile
\and Departamento de F\'isica, Facultad de Ciencias B\'asicas, Universidad Metropolitana de Ciencias de la Educaci\'on, Av. Jos\'e Pedro Alessandri 774, \~Nu\~noa, Santiago, Chile
\and {Centro de Desarrollo de Investigaci\'on UMCE, Universidad Metropolitana de Ciencias de la Educaci\'on, Santiago, Chile}
\and Institut de Recherche en Astrophysique et Planétologie (IRAP), UPS-OMP, CNRS, CNES, 9 avenue du Colonel Roche, BP 44346, F-31028 Toulouse Cedex 4, France
\and Department of Astronomy, Indiana University, Bloomington, IN 47405, USA
}

\date{Received XX Month 2026 / Accepted XX Month 2026}


\abstract
{The detection of likely thermal ultraviolet emission from a few old neutron stars suggests that at least one internal heating mechanism is present in these stars. One proposed mechanism is rotochemical heating, in which the continuous contraction of the neutron star due to its spin-down produces chemical imbalances that induce Urca reactions, and the latter deposit heat in the neutron star core. If the protons in the star are superconducting, their energy gap suppresses the reactions, except in microscopic magnetized regions (such as quantized flux tubes) in which the protons act as if they were normal. Therefore, the strength of the internal magnetic field controls the rate at which reactions proceed and thus affects the thermal evolution of the neutron star.
}
{Here, we present the first comprehensive study of the effect of an internal magnetic field in the superconducting interior on rotochemical heating.}
{We simulate the evolution of neutron stars for different internal magnetic field strengths and neutron energy gaps, comparing the results to Hubble Space Telescope observations of old neutron stars.}
{All the observational data can be accounted for if the proton energy gap is large ($\sim 1.5\,\mathrm{MeV}$) and the neutron energy gap is small ($\lesssim 0.1\,\mathrm{MeV}$) or vanishing, while the millisecond pulsar PSR~J0437$-$4715 needs to have a very weak internal magnetic field.}
{Our results suggest that neutron-star cores are characterized by a large proton pairing gap and a small or vanishing neutron gap, and that millisecond pulsars have very weak internal magnetic fields. Under these conditions, rotochemical heating alone can account for the observed thermal emission of old neutron stars.
}


\maketitle


\section{Introduction}\label{sec:intro}

The thermal history of neutron stars (NSs) depends on the properties of superdense matter, such as its composition, superfluidity, and superconductivity, allowing to probe these properties observationally \citep{Be2018}. Passive cooling models predict that the surface temperature will become undetectably low within $\sim 10^6\,\mathrm{yr}$ \citep{Yako2004}. However, Hubble Space Telescope (HST) observations have found several old pulsars emitting potentially thermal ultraviolet radiation, namely
the Gyr-old millisecond pulsars (MSPs) PSR~J0437$-$4715 (hereafter J0437; \citealt{Kar2004,Durant2012}) and PSR~J2124$-$3358 (hereafter J2124; \citealt{Range2017}), and 
the $\sim 10^{7}$ and $\sim 10^{8}$ yr old classical pulsars (CPs) PSR~B0950+08 (hereafter B0950; \citealt{Pavlov2017,Abramkin2022}) and  PSR~J0108$-$1431 (hereafter J0108; \citealt{Abramkin2021}), respectively.
Thus, internal heating mechanisms appear to operate in these stars. On the other hand, the old CP PSR~J2144$-$3933 (hereafter J2144) has not been detected \citep{Guillot2019}, making it the coolest NS known and providing a constraint on the potential heating mechanisms.

Some of the proposed heating mechanisms are rotochemical heating \citep{Reis1995}, vortex creep \citep{Alpar1984}, dark matter accretion \citep{Kou10,deLava}, crust cracking \citep{Baym71,Cheng1992}, and magnetic field decay \citep{Gold92,Thom96,Pons2025}. \citet{GonRei2010} established that only rotochemical heating and vortex creep might explain the inferred surface temperature of J0437, and \citet{Rodriguez2025} (hereafter R25) found that a model combining these two mechanisms can explain all the observations mentioned above. 

The present work focuses on rotochemical heating, which operates as follows \citep{Reis1995}. The NS slowly decreases its rotation rate and thus its centrifugal force, progressively increasing the density of all matter elements in its interior. This changes the chemical potentials $\mu_i$ of different particle species by different amounts, causing a chemical imbalance, e.~g., $\eta_\ell\equiv\mu_n-\mu_p-\mu_\ell$, where $n, p, \ell$ denote neutrons, protons, and leptons (electrons [$\ell=e$] or muons [$\ell=\mu$]), respectively. Thus, non-equilibrium reactions occur in the NS core, which emit neutrinos and deposit heat. 

If the neutrons are superfluid, the protons are superconducting, or both, there will be non-zero energy gaps, $\Delta_n$, $\Delta_p$, around the respective Fermi surfaces, and the reactions will be strongly suppressed as long as the chemical imbalances are smaller than a certain threshold $\Delta_{\mathrm{thr}}$. Once the threshold is crossed, reactions turn on quickly, with two possible outcomes: a quasi-steady state with nearly constant chemical imbalances and temperature (\citealt{Reis1997,Petro2010,GonJ2015}; R25) or strong oscillations of the same variables (\citealt{Petro2011}; R25). The latter are only seen in cases where the gaps are large and uniform across the NS core, which is not realistic, motivating R25 to reduce the reaction rates by a certain factor in order to mimic the realistic case of a non-uniform gap and suppress the oscillations. 

None of the works cited in the previous paragraph considered that, if the proton superconductor is threaded by a magnetic field, the latter is confined to regions in which the protons behave as ``normal'' (non-superconducting) particles (quantized flux tubes in a type-II superconductor and irregular domains in a type-I superconductor), allowing for faster reactions in these regions \citep{Schaab1998,Kantor2021}. These regions of normal protons occupy a fraction $f\approx B_{\rm int}/H_{\rm crit}$ of the NS core, where $B_{\rm int}$ is the average magnetic flux density in this region and $H_{\rm crit}$ is a critical magnetic field of the superconductor (the upper critical field $H_{\rm c2}$ for type II, the thermodynamic critical field $H_{\rm c}$ for type I). In the present work, we consider this as an additional degree of freedom and explore whether the thermal emission of old NSs can be explained by rotochemical heating with this effect.

Sect.~\ref{sec:model} presents the physical model used in this work. Sect.~\ref{sec:Setup} describes the general setup of our simulations, and Sect.~\ref{sec:Results} describes their results. Finally, Sect.~\ref{sec:summary} summarizes and discusses the main results.

\section{Model}
\label{sec:model}

In this section, we briefly discuss the main equations needed to understand the present work. More details can be found in \citet{Petro2010} and in R25.

We will consider cooling and rotochemical heating of a NS core composed of degenerate neutrons ($n$), protons ($p$), electrons ($e$), and muons ($\mu$). At high enough densities (not necessarily reached in NSs), these particles can convert into each other by direct Urca (hereafter Durca) reactions,
\begin{equation}\label{dur}
n \rightarrow p+\ell+\bar\nu_\ell,\qquad p +\ell \rightarrow n+ \nu_\ell,
\end{equation}
whereas at lower densities only the less efficient modified Urca (Murca) reactions are allowed,
\begin{equation}\label{mur}
n + N \rightarrow p+\ell+\bar\nu_\ell + N,\qquad p +\ell + N \rightarrow n+ \nu_\ell + N.
\end{equation}
Here, $\ell$ is a negatively charged lepton (electron or muon), $\nu_\ell$ and $\bar\nu_\ell$ are the corresponding neutrino or antineutrino, and $N$ is a bystander nucleon (neutron or proton) that exchanges momentum without changing its flavor.

Since we are interested in timescales much longer than the diffusion times, we can consider the redshifted core temperature, $T_{\rm c}^\infty$, and the redshifted chemical potentials, $\mu_i^\infty$, to be uniform throughout the NS core \citep{Reis1995}. (Throughout this paper, a subscript or superscript $\infty$ denotes a quantity as measured by a distant observer.) The evolution of the core temperature is given by \citep{Thorne1977}
\begin{equation} \label{ThermalEvol}
\dot T_{\rm c}^\infty = \frac{1}{C}\Big(L_{\rm H}^{\infty}-L_{\nu}^{\infty}-L_{\gamma}^{\infty}\Big),
\end{equation}
where $C$ is the total heat capacity of the star, $L_{\rm H}^{\infty}$ is the total power injected by rotochemical heating, $L_{\nu}^{\infty}$ is the power emitted as thermal neutrinos and antineutrinos from the NS interior, and
\begin{equation} \label{photon luminosity}
    L_{\gamma}^{\infty}=4\pi R_\infty^2\sigma (T^\infty_{\rm s})^4
\end{equation} 
is the power released as thermal photons from the surface. In the latter equation, $\sigma$ is the Stefan-Boltzmann constant, $R_\infty$ is the NS radius, and $T_{\rm s}^\infty$ is its (effective) surface temperature. Throughout this paper, dots denote time-derivatives. The temperatures of the core, $T_{\rm c}^\infty$, and the surface, $T_{\rm s}^\infty$, are in a direct relation that depends on the envelope composition \citep{Gud83,Potekhin1997}. 

Both $L_{\rm H}^{\infty}$ and $L_{\nu}^{\infty}$ are due to Urca processes, and the former can be written as
\begin{equation} \label{eq:heating}
    L_{\rm H}^{\infty}=\sum_{\ell=e,\mu}\eta_\ell^\infty \Delta\tilde\Gamma^\infty_{np\ell},
\end{equation} 
where $\Delta\tilde\Gamma^\infty_{np\ell}$ is the net rate at which neutrons are converted into protons and electrons ($\ell=e$) or protons and muons ($\ell=\mu$). 
These rates are functions of the chemical imbalances, the temperature, and, if applicable, the Cooper pairing gaps of neutrons, $\Delta_n$, and/or protons, $\Delta_p$ \citep{Petro2010}. In particular, when $k_BT_{\rm c}$ is small compared to the chemical imbalances and the energy gaps, the rates are strong functions of the chemical imbalances. ($k_B$ is the Boltzmann constant.) They are very small for small $\eta_\ell$ and increase sharply for $\eta_\ell\approx\Delta_{\mathrm{thr}}$, where the latter is given by
\begin{equation}\label{thresh}
\Delta_{\text{thr}}=\begin{cases}
\Delta_n+\Delta_p, & \text{for Durca reactions,} \\
\min\{\Delta_n + 3\Delta_p, 3 \Delta_n + \Delta_p\} & \text{for Murca reactions.} 
\end{cases}
\end{equation}
For $\eta_\ell$ much larger than both $k_BT_{\rm c}$ and $\Delta_{\rm thr}$ (including the non-Cooper-paired case with $\Delta_{\rm thr}=0$), the rates satisfy $\Delta\tilde\Gamma_{np\ell}\approx A_\ell\eta_\ell^n$, where $A_\ell$ is a constant and $n=5$ for Durca and $n=7$ for Murca reactions.

The evolution of the chemical imbalances is given by
\begin{eqnarray}\label{eq:deta/dt}
    \dot\eta_e^\infty=2W_{npe}\Omega\dot\Omega-Z_{npe}\Delta\tilde\Gamma^\infty_{npe}-Z_{np}\Delta\tilde\Gamma^\infty_{np\mu}, \\
    \dot\eta_\mu^\infty=2W_{np\mu}\Omega\dot\Omega-Z_{np}\Delta\tilde\Gamma^\infty_{npe}-Z_{np\mu}\Delta\tilde\Gamma^\infty_{np\mu},  
\end{eqnarray}
where $\Omega$ is the angular velocity of rotation of the NS, 
and $W_{npe}$, $W_{np\mu}$, $Z_{npe}$, $Z_{np\mu}$, and $Z_{np}$ are constants that depend on the structure of the NS \citep{Fer2005,Reisenegger2006,Petro2010}. The first term on the right-hand side of both equations corresponds to the increase of the chemical imbalances due to compression of the matter as the star spins down, whereas the remaining terms account for their decrease due to reactions\footnote{Note that $W_{npe}<0$ and $W_{np\mu}<0$, therefore, when the star spins down ($\dot\Omega<0$), it makes the chemical imbalances increase. On the other hand, $Z_{npe}$, $Z_{np\mu}$, and $Z_{np}$ are all $>0$, and $\Delta\tilde\Gamma^\infty_{np\ell}$ has the same sign as $\eta_\ell$ for $\ell=e,\mu$, so the terms involving these quantities always make $|\eta_\ell|$ decrease.}.

\subsection{Rotochemical heating with a magnetized proton superconductor} 
\label{sec:rotochemical with flux tubes}

An effect not considered in most of the work on rotochemical heating so far, but initially pointed out by \citet{Schaab1998} and later discussed and evaluated by \citet{Kantor2021}, is that, when the protons undergo the transition to the superconducting state, the magnetic flux gets concentrated into small regions in which the protons continue to behave as normal particles. In the outer part of the core, protons are expected to be a type-II superconductor, in which these regions are quantized flux tubes with normal cores, whereas, above some threshold density, the superconductor might be type I, with the magnetic flux confined to irregular domains containing normal protons \citep{Sedrakian2005}.

In both cases, the volume fraction of the protons that remain normal is {$f\approx B_{\rm int}/H_{\rm crit}$}, where $B_{\rm int}$ is the average magnetic flux density inside the core, and $H_{\rm crit}$ is a critical magnetic field, namely the upper critical field, $H_{\rm c2}$, for a type II superconductor, and the thermodynamic critical field, $H_{\rm c}$, for a type I superconductor \citep{Tilley1990}.\footnote{When doing numerical evaluations in this paper, we take $H_{\rm crit}=10^{15}\,\mathrm{G}$ as an approximation to the uncertain true values \citep{Kantor2021}.} Thus, for $\eta_\ell, k_BT_{\rm c}\ll\Delta_p$, Urca reactions are strongly suppressed, except possibly within this small magnetized volume \citep{Schaab1998,Kantor2021}. 
In this model, there will be two different thresholds for the chemical imbalances, with a small (or vanishing) threshold $\Delta_{\rm thr}^{\rm in}=\Delta_n$ in the magnetized regions, and Eq.~(\ref{thresh}) still applying for the threshold $\Delta_{\rm thr}^{\rm out}$ in the bulk, unmagnetized region. This means that, as the chemical imbalances increase, significant reactions turn on first in the magnetized regions, and only later, if at all, in the unmagnetized regions. 

In principle, the same 
occurs in the cores of the quantized vortices {due to} 
the rotation of the neutron superfluid. However, these vortices are much fewer and occupy a much smaller volume than the magnetized regions of the proton superconductor; therefore, we ignore their effect.

\begin{table}
\begin{center}
\begin{tabular}[H]{lcccc}
\hline

Pulsar & 
$P$ [ms]       & 
log $B$ [G] & 
log  $\tau$ [yr] &
$T^\infty_{\rm s}/(10^5 \mathrm{K})$\\
\hline

J0437$-$4715  & 
5.75          & 
8.45          & 
9.82          & 
$2.36^{+0.15}_{-0.14}$\\ 

J2124$-$3358 & 
4.93         &
8.28         &
10.02        & 
$< 1.8$\\
\hline

B0950$+$08 & 
253      & 
11.38    & 
7.24     & 
$0.6-1.2$\\ 

J0108$-$1431 & 
808      & 
11.36    & 
8.29     & 
$<0.6$\\ 

J2144$-$3933 & 
8510         &
12.27        &
8.43         & 
$<0.3$\\ 
\hline
\end{tabular}
\caption{Properties of the pulsars considered in the present paper (from R25). The columns contain the name of the pulsar, its rotation period $P$, magnetic field strength $B=3.2\times 10^{19}(P[s]\dot P)^{1/2}$ inferred from spin-down, characteristic age $\tau=P/(2\dot P)$, and measured surface redshifted temperature ($T_{\rm s}^\infty$). The values of $P$, $B$, and $\tau$ were taken from the ATNF Pulsar Catalogue ($B$ and $\tau$ are corrected for the ``Shklovskii effect'', \citealt{Shklovskii1970,Camilo1994}), and $T_{\rm s}^\infty$ from the HST observations of J0437 \citep{Gon2019}, J2124 \citep{Range2017}, B0950 \citep{Abramkin2022}, J0108 \citep{Abramkin2021}, and J2144 \citep{Guillot2019}, applying recalibrations for J2124 and J2144 (see R25). We note that the first two pulsars are MSPs, whereas the last three are CPs.
}\label{table:summary}
\end{center}
\end{table}

\section{General setup of the simulations} 
\label{sec:Setup}

We calculate thermal evolution curves for different parameters using the code of \citet{Petro2010} as modified by R25, and compare the results to the same observational data used and described by R25 and summarized in Table~\ref{table:summary}. We note that the pulsars considered fall into two groups: The first two are millisecond pulsars (MSPs) with short rotation periods, weak magnetic fields, and long characteristic ages, and the remaining three are classical pulsars (CPs) with the opposite properties. 
As in R25, the different dipole field strengths of MSPs and CPs make them behave in qualitatively different ways, which we therefore discuss separately. We run simulations corresponding to the spin-down parameters of the pulsars with reliable temperature measurements (as opposed to upper limits), namely J0437 as representative of MSPs and B0950 of CPs, comparing also to the upper limit for J0108, whose inferred dipole field is very similar to that of B0950. We verified that the upper limits on J2124 and J2144 do not change our conclusions.

As in R25, we always assume magnetic dipole spin-down with a constant magnetic field $B=3.2\times 10^{19}(P[s]\dot P)^{1/2}$ G inferred from the observed values of the period $P$ and its time-derivative $\dot P$ for the respective pulsar. We set the initial period to $P_0=1\,\mathrm{ms}$ for MSPs and $P_0=5$ ms for CPs. The age of each pulsar is taken to be the time $t$ at which the computed spin-down parameters $P(t)$ and $\dot P(t)$ match the observed values. This time is generally slightly shorter than the characteristic age, $\tau\equiv P/(2\dot P)$, which would be the same in the limit of infinitely rapid initial rotation. 

{In order to relate $T_{\rm c}^\infty$ to $T_{\rm s}^\infty$, for CPs we assume a fully catalyzed envelope containing heavy elements \citep{Gud83}, whereas for MSPs we assume an accreted envelope with light elements \citep{Potekhin1997}, but we note that the late-time evolution of $T_{\rm s}^\infty$ does not depend on this choice (R25).} We assume an initial core temperature $T_{\rm c,0}^\infty=10^{11}\,\mathrm{K}$ for CPs (as expected from core collapse) and $T_{\rm c,0}^\infty=10^9\,\mathrm{K}$ for MSPs (corresponding to the end of their accretion phase), and initial chemical imbalances $\eta_{\ell,0}^\infty=0$. In practice, these values do not affect the late-time evolution we are interested in (see, e.~g., \citealt{Reis1995,Fer2005,Kantor2021}). 

We use a NS model of mass $M=1.44\,M_\odot$ and equation of state A18$+\delta v+$UIX$^*$, containing neutrons, protons, electrons, and muons. This is Model M in R25, where its main parameters can be found. It does not allow direct Urca reactions to occur, so modified Urca reactions are the main mechanism for the NS to emit neutrinos and to adjust its composition. In all simulations, we assume a large, uniform, and isotropic proton gap $\Delta_p^\infty=1.5\,\mathrm{MeV}$, as we argue that such a value is required to fit the observation of J0437. On the other hand, we treat the (also uniform and isotropic) neutron gap $\Delta_n^\infty$ and the magnetized volume fraction $f\equiv B_{\rm int}/H_{\rm crit}$ as free parameters that we vary between simulations.

\begin{figure}

\centering
    \includegraphics[width=0.5\linewidth]{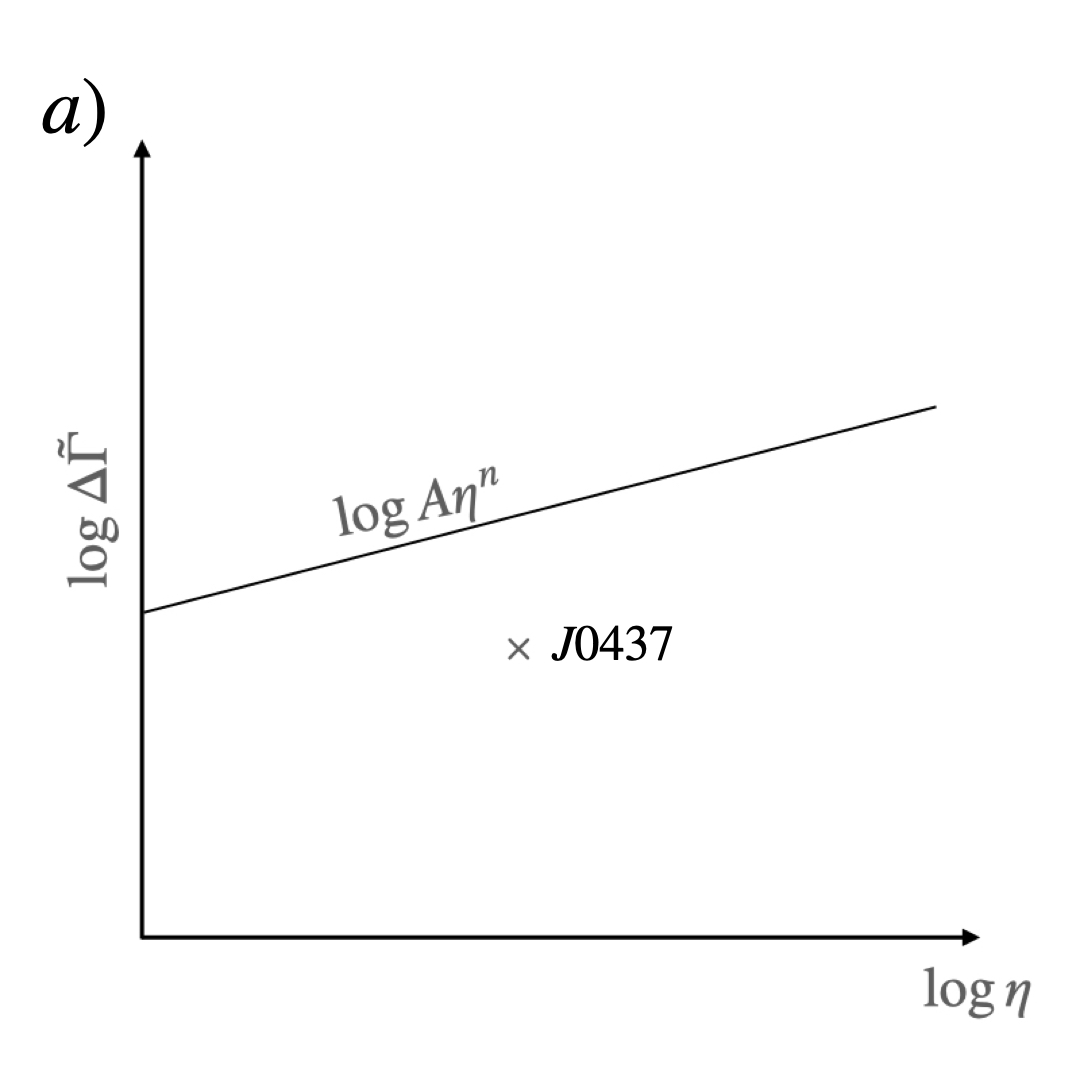}\hfil
    \includegraphics[width=0.5\linewidth]{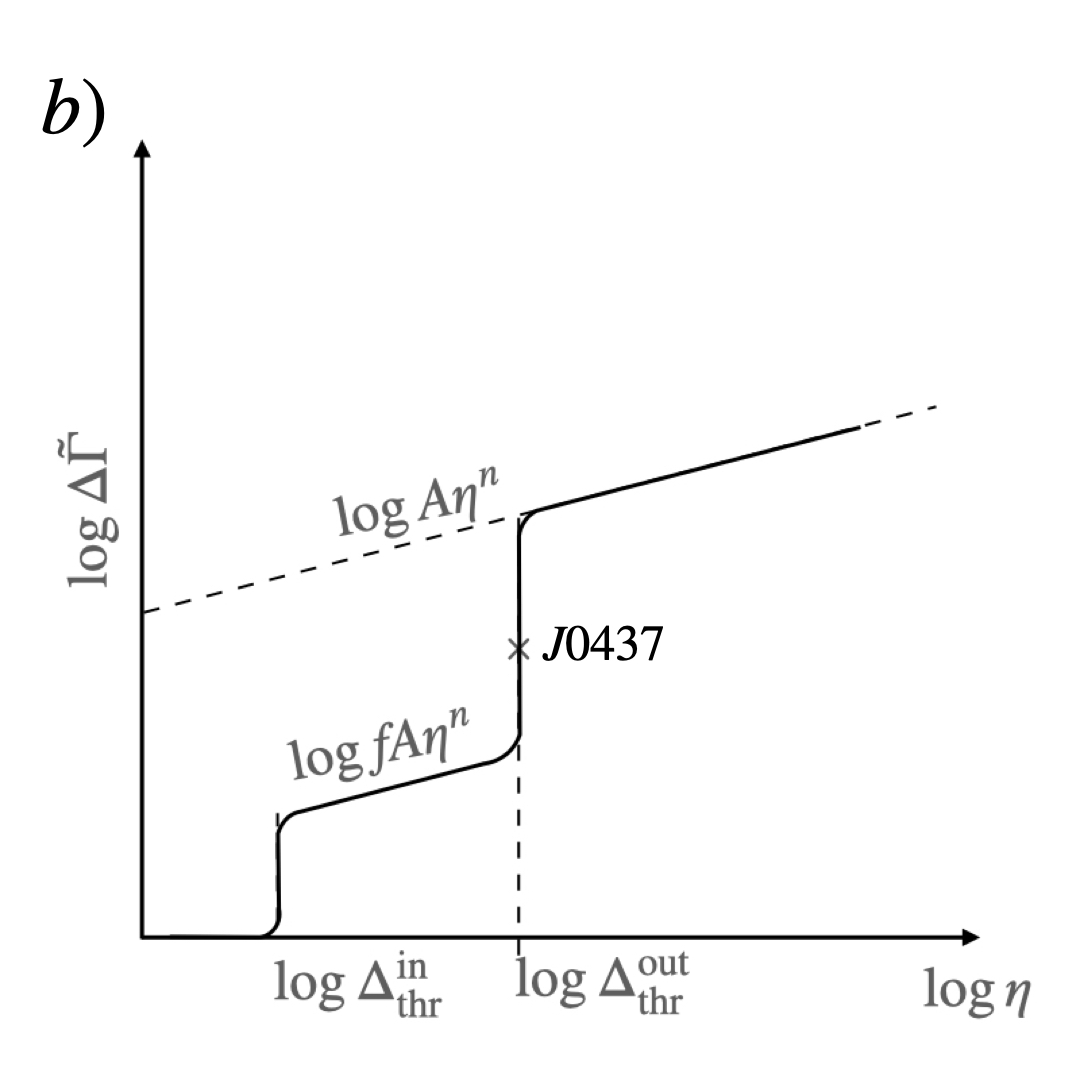}\par\medskip
    \includegraphics[width=0.5\linewidth]{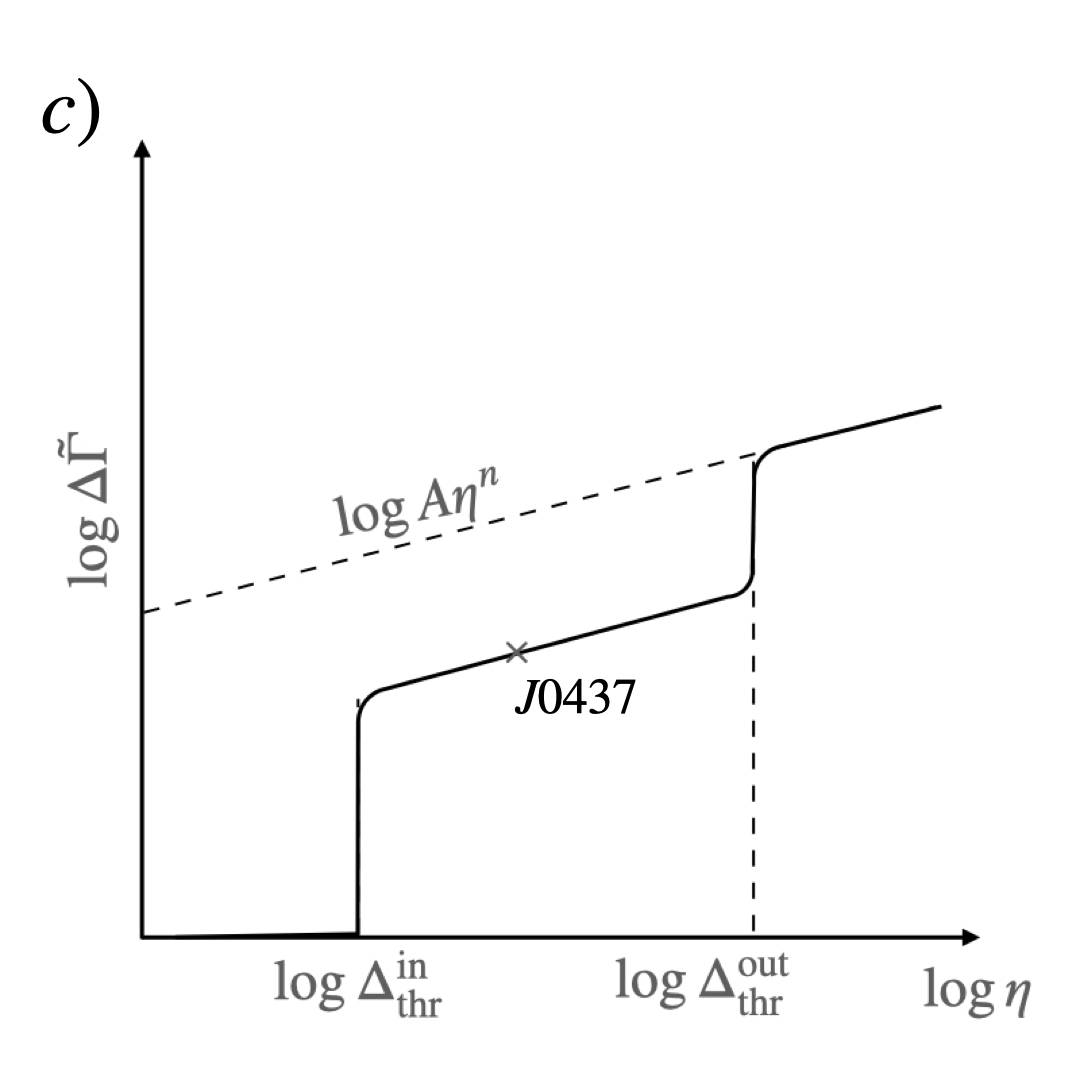}\hfil
    \includegraphics[width=0.5\linewidth]{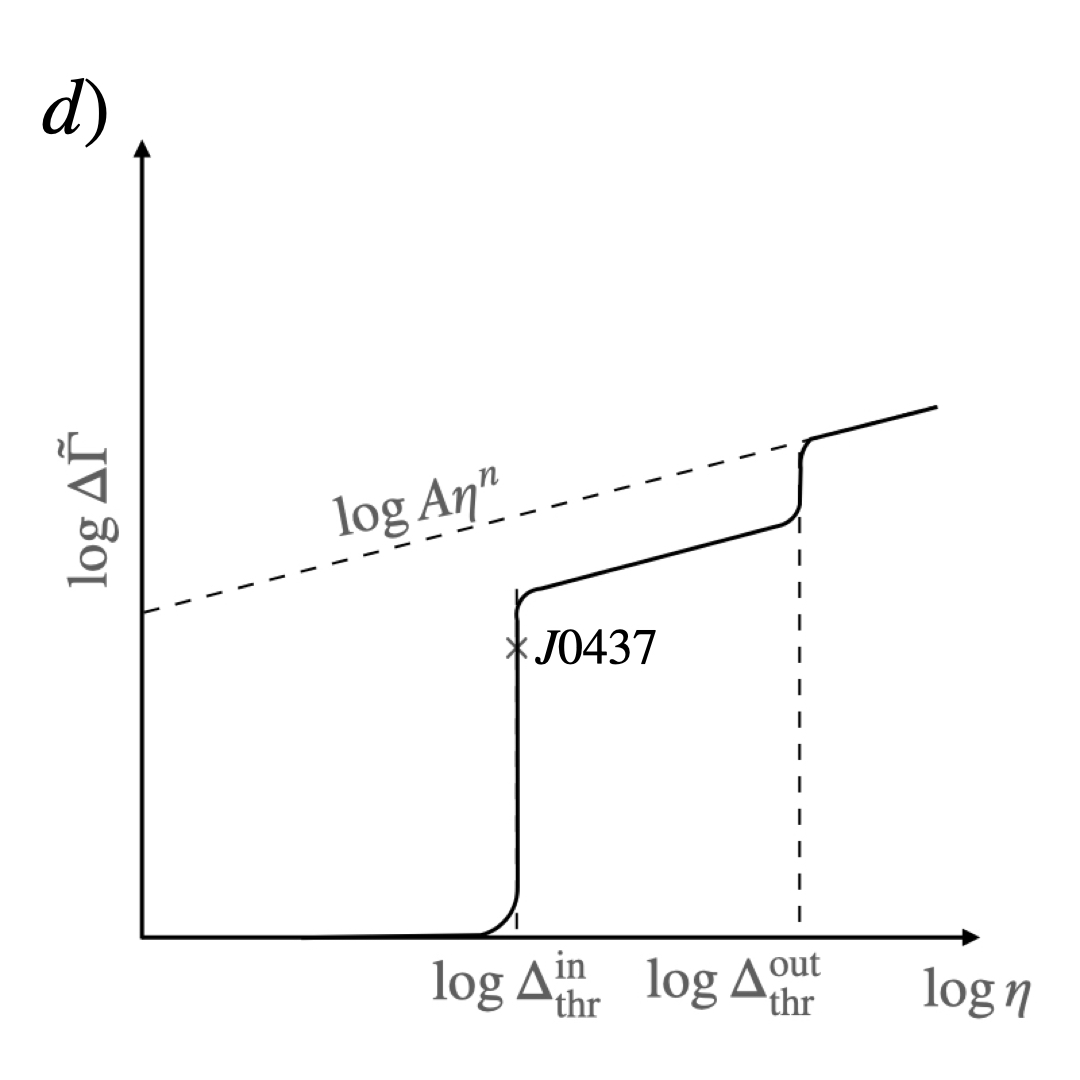}\hfil

\caption{Schematic representation of the net reaction rate $\Delta\tilde\Gamma_{np\ell}^\infty$ as a function of the chemical imbalance $\eta_\ell^\infty$, for four different cases: (a) Normal neutrons and protons; (b) $\Delta_{\rm thr}^{\rm in,\infty}<\eta_\ell^\infty\approx\Delta_{\rm thr}^{\rm out,\infty}$ and $f\lesssim f^*$; (c)~$\Delta_{\rm thr}^{\rm in,\infty}\lesssim\eta_\ell^\infty\lesssim\Delta_{\rm thr}^{\rm out,\infty}$ and $f\approx f^*$; (d) $\Delta_{\rm thr}^{\rm in,\infty}\approx\eta_\ell^\infty<\Delta_{\rm thr}^{\rm out,\infty}$ and $f\gtrsim f^*$.
}\label{schematic}
\end{figure}

\section{Results}
\label{sec:Results}

\subsection{Millisecond pulsars}
\label{sec:MSPs with flux tubes}

It has been shown that, under diverse circumstances, MSPs with rotochemical heating tend to reach a quasi-steady state in which the chemical imbalances and temperature remain approximately constant \citep{Reis1995,Reis1997,Fer2005,Petro2010,GonJ2015,Yanagi2020,Kopp2023}. In this state, the reduction of the chemical potentials through Urca reactions compensates for their increase due to the spin-down, 
\begin{equation}
    \Delta\tilde\Gamma_{np\ell}^\infty\approx-K_\ell\Omega\dot\Omega,
\end{equation}
and all the energy released into the stellar interior is emitted as thermal photons from the surface,
\begin{equation}
    L_\gamma^\infty\approx\sum_{\ell=e,\mu}h_\ell\eta_\ell^\infty\Delta\tilde\Gamma_{np\ell}^\infty\approx-\Omega\dot\Omega\sum_{\ell=e,\mu}h_\ell K_\ell\eta_\ell^\infty,
\end{equation}
where $K_\ell$ are constants that depend on the stellar structure and $h_\ell$ are dimensionless coefficients of order unity (see R25). Thus, if the spin-down parameters $\Omega$ and $\dot\Omega$ as well as the photon luminosity $L_\gamma^\infty$ for a certain MSP have been measured, the net reaction rates $\Delta\tilde\Gamma_{np\ell}^\infty$ and the ``average'' chemical imbalance
\begin{equation}
    \bar\eta_\infty\equiv\frac{1}{K}\sum_{\ell=e,\mu}h_\ell K_\ell \eta_\ell^\infty\approx\frac{L_\gamma^\infty}{-K\Omega\dot\Omega},
\end{equation}
where $K\equiv K_e+K_\mu$, can be inferred. For our NS model with the parameters of J0437, we obtain $\bar\eta^\infty\approx 1.5\,\mathrm{MeV}$. In the approximate arguments that follow, we assume that the relevant parameters of electrons and muons are the same, so we make no distinction between them.

In the absence of Cooper pairing ($\Delta_n=\Delta_p=0$) and for $\eta_\ell^\infty\gg k_BT_{\rm c}^\infty$, the net reaction rates have the form $\Delta\tilde\Gamma_{np\ell}^\infty\approx A_\ell(\eta_\ell^\infty)^n$, where $n=5$ for Durca and $n=7$ for Murca reactions, and $A_\ell$ are constants (for $\ell=e,\mu$). For J0437, this line lies above the point marking the pulsar in Fig.~\ref{schematic}, panel~(a), therefore the latter cannot be explained with normal matter, and Cooper pairing needs to be invoked. 

In the most general case with finite $\Delta_n$ and $\Delta_p$ and with a magnetized fraction $f$ in the proton superconductor, 
the reaction rates increase strongly as the chemical potentials cross the two threshold values, 
$\Delta_{\rm thr}^{\rm in}=\Delta_n$ in the magnetized regions and 
$\Delta_{\rm thr}^{\rm out}$ (given by Eq.~\ref{thresh}) in the rest of the superconductor. Specifically, the net reaction rates behave as follows as a function of the chemical imbalances: 
\begin{equation}\label{alternatives}
\Delta\tilde\Gamma_{np\ell}^\infty\approx\begin{cases}
0, & \text{for } \eta_\ell^\infty\lesssim\Delta_{\rm thr}^{\rm in,\infty}, \\
fA_\ell(\eta_\ell^\infty)^n & \text{for }\Delta_{\rm thr}^{\rm in,\infty}\lesssim\eta_\ell^\infty\lesssim\Delta_{\rm thr}^{\rm out,\infty}, \\
A_\ell(\eta_\ell^\infty)^n  & \text{for }\eta_\ell^\infty\gtrsim\Delta_{\rm thr}^{\rm out,\infty}.
\end{cases}
\end{equation}
Therefore, as illustrated by panels (b), (c), (d) of Fig.~\ref{schematic}, there are three qualitatively different possibilities to fit the data point of J0437:
\begin{itemize}
    \item In panel~(b), $\Delta_{\rm thr}^{\rm in,\infty}<\eta_\ell^\infty\approx\Delta_{\rm thr}^{\rm out,\infty}$ and $f\lesssim f^*$,
    \item in panel~(c), $\Delta_{\rm thr}^{\rm in,\infty}\lesssim\eta_\ell^\infty\lesssim\Delta_{\rm thr}^{\rm out,\infty}$ and $f\approx f^*$, 
    \item in panel~(d), $\Delta_{\rm thr}^{\rm in,\infty}\approx\eta_\ell^\infty<\Delta_{\rm thr}^{\rm out,\infty}$ and $f\gtrsim f^*$,
\end{itemize} 
where 
\begin{equation}
    f^*\approx{(K_\ell\Omega|\dot\Omega|)^{n+1}\over A_\ell(L_\gamma^\infty)^n}. 
\end{equation}
For our NS model with the data for J0437 (assuming Murca reactions only), we obtain $f^*\approx 6\times 10^{-14}$, with a high uncertainty due to the large exponents and the uncertain values of both the theoretical and observational input parameters. We note that all three possibilities require $\Delta_{\rm thr}^{\rm out,\infty}\gtrsim\eta_\ell^\infty\approx 1.5\,\mathrm{MeV}$, thus at least one of the gaps must be relatively large.

\begin{figure}
\includegraphics[width=1\linewidth]{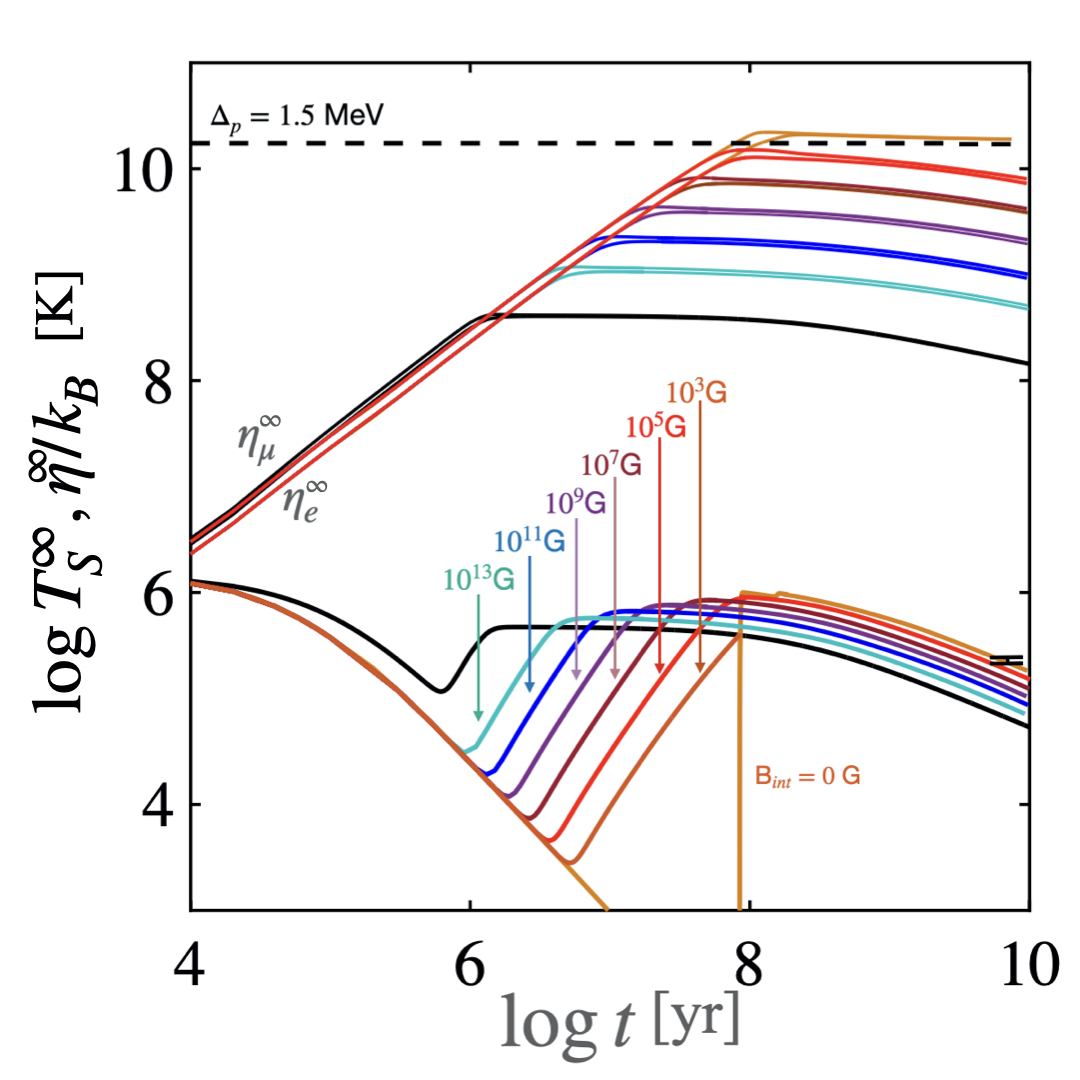}
\caption{Evolution of the redshifted surface temperature ($T_{\rm s}^\infty$) and electron ($\eta_e^\infty$) and muon ($\eta_\mu^\infty$) chemical imbalances for rotochemical heating due to spin-down with the inferred magnetic dipole moment of PSR~J0437-4715, assuming only Murca reactions are allowed. The temperature measurement for this pulsar is given by the error bar, placed at the time when the spin parameters $P$ and $\dot P$ reach the measured values. The solid black lines correspond to normal neutrons and protons. The solid lines in cyan, blue, purple, brown, red, chocolate, and orange consider normal neutrons and superconducting protons with a uniform Cooper pairing gap, $\Delta_p=1.5$ MeV, core magnetic flux densities $B_{\rm int}=10^{13}, 10^{11}, 10^{9}, 10^7, 10^5, 10^3$, and 0 G, respectively, assuming a critical magnetic field $H_{\rm crit}=10^{15}$ G. For each colour, from top to bottom, there are three curves representing $\eta_\mu^\infty/k_B$, $\eta_e^\infty/k_B$, and $T_{\rm s}^\infty$, where $k_B$ is the Boltzmann constant. All curves were computed considering a NS with an initial rotation period $P_0=1$~ms, initial core temperature $T_0^\infty=10^9$~K, equation of state A18+$\delta v$+UIX$^*$ \citep{Akmal1998}, mass $M=1.44\,M_{\odot}$, and coordinate radius $R=11.45\,\mathrm{km}$. 
}\label{Bint}
\end{figure}

\subsubsection{Superconducting protons and normal neutrons}
\label{sec:MSP normal neutrons}

A particular case of the scenario illustrated by panel~(b) of Fig.~\ref{schematic} is to take the neutrons to be normal ($\Delta_{\rm thr}^{\rm in}=\Delta_n=0$). We do this in Fig.~\ref{Bint}, where we contrast the effect of different values of $f$ for MSPs with normal neutrons 
and superconducting protons with a large energy gap ($\Delta_p^\infty=1.5\,\mathrm{MeV}=\Delta_{\rm thr}^{\rm out,\infty}$) {and considering only Murca reactions}, confirming the results shown in Fig.~3 of \citet{Kantor2021}. We note that the behavior is always qualitatively similar to the case of normal protons and neutrons, with the energy gaps and temperatures stabilizing in a quasi-steady state. 

For large $f$, 
the temperature starts to increase at an early time, and both the chemical imbalances and the temperature saturate at relatively low values. In this regime, $\Delta\tilde\Gamma_{np\ell}^\infty\approx fA_\ell(\eta_\ell^\infty)^n$, and therefore the quasi-steady values are $\eta_\ell^\infty\approx(K_\ell\Omega|\dot\Omega|/[fA_\ell])^{1/n}$ and $L_\gamma^\infty\approx\sum_\ell h_\ell (fA_\ell)^{-1/n}(K_\ell\Omega|\dot\Omega|)^{1+1/n}$, insufficient to match the observation of J0437. 
In the extreme case of $f=1$, all protons become normal, recovering the case studied by \citet{Fer2005}. 

On the other hand, for extremely small $f$ {(here $f\lesssim 10^{-12}$)}, the effective volume for reactions to occur becomes {so small that} 
the case of an unmagnetized proton superconductor {is approached}, in which the chemical imbalances grow unimpeded until $\eta_\ell^\infty\approx\Delta_{\text{thr}}^{\rm out,\infty}=\Delta_p^\infty$), when reactions {in the bulk superconductor} turn on rather suddenly, releasing a substantial amount of energy, and thus keeping the imbalances and the temperature at higher values than {in the case of normal matter \citep{Reis1997,Petro2010}. For our model and taking the measured temperature of J0437 at face value, only such extremely low values of $f$ can fit the observation, and only for a large proton gap, $\Delta_p\gtrsim 1.5\,\mathrm{MeV}$.} We verified that, for these parameters, the upper limit for J2124 is also satisfied, though not by a large margin.

\begin{figure}
\includegraphics[width=1\linewidth]{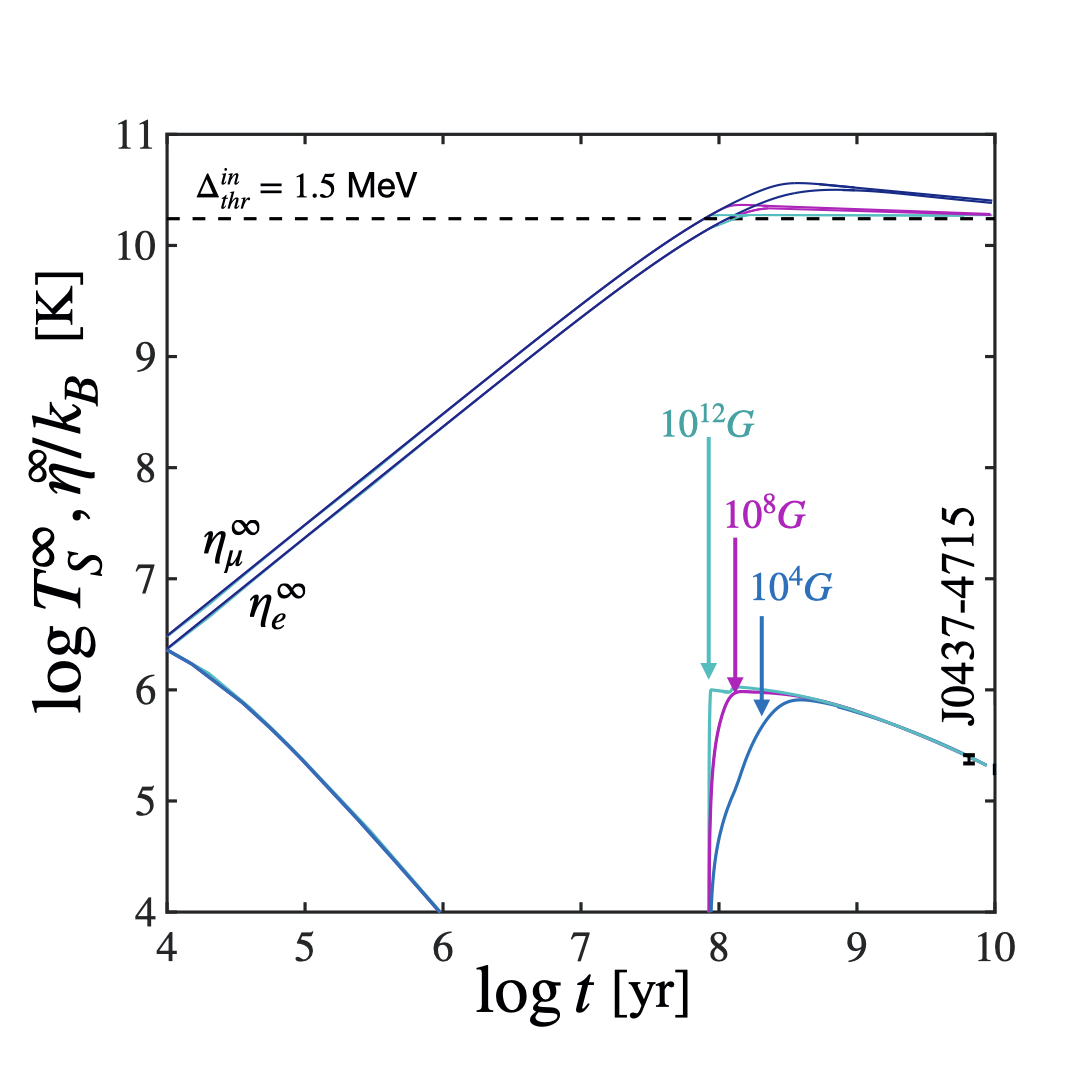}
\caption{The same as Fig.~\ref{Bint}, now for superconducting protons and superfluid neutrons with $\Delta_p=\Delta_n=1.5$ MeV, for different internal magnetic fields $B_{\rm int}=10^{12}$ G (cyan),  $10^{8}$ G (purple) and $10^{4}$ G (blue), assuming $H_{\rm crit}=10^{15}\,\mathrm{G}$}. 
\label{BintneutronMSP}
\end{figure}

\subsubsection{Superconducting protons and superfluid neutrons}
\label{sec:MSP superfluid neutrons}

Now, we consider the case of Fig.~\ref{schematic}, panel (d), with superconducting protons as well as superfluid neutrons with a large gap, $\Delta_n^\infty=1.5\,\mathrm{MeV}$ (see Fig.~\ref{BintneutronMSP}). 
For $|\eta_\ell^\infty|\ll\Delta_{\text{thr}}^{\rm in,\infty}=\Delta_n^\infty$, 
reactions are strongly suppressed everywhere, {so the chemical imbalances grow at least until $\eta_\ell^\infty\approx
\Delta_n^\infty$, at which point 
reactions are activated in the magnetized regions. For very large $f\gtrsim 10^{-2}$, this causes thermal oscillations, as observed by \citet{Petro2011} and R25. For somewhat smaller $f$ in the range $f^*\lesssim f\lesssim 10^{-2}$, the activation of the reactions prevents a further increase of the chemical imbalances and causes a heat release} that keeps the star at a relatively high quasi-steady temperature, nearly the same as if only the neutrons were superfluid. 
{In this regime, the luminosity can be approximated by the expression found by \citet{Petro2010} for the case of superfluid neutrons and normal protons},
\begin{equation}
L_\gamma^\infty \approx (1-4)\times 10^{32}\frac{\Delta_n^\infty}{\text{MeV}}\frac{\dot P_{-20}}{P_\mathrm{ms}^3}\text{erg s}^{-1},
\end{equation} 
where $P_\mathrm{ms}$ is the rotation period in milliseconds and $\dot P_{-20}$ its time-derivative in units of $10^{-20}$. This relation is valid for Murca and Durca reactions and regardless of the values of $\Delta_p^\infty$ and $f$ in the range $f^*\lesssim f\lesssim 10^{-2}$. 
This is illustrated in Fig.~\ref{BintneutronMSP} for large energy gaps of both protons and neutrons, $\Delta_p^\infty=\Delta_n^\infty=1.5\,\mathrm{MeV}$, and a few very different values of $f$. 
For all of these, the chemical imbalances grow until they reach $\eta_\ell^\infty\approx\Delta_n^\infty$, yielding a nearly identical temperature evolution at all later times. Here, the value of $\Delta_n^\infty=1.5\,\mathrm{MeV}$ was chosen 
so as to match the temperature of J0437, also satisfying the upper limit for J2124.

If $\Delta_n^\infty$ is smaller than assumed in Fig.~\ref{BintneutronMSP}, the only way the observed temperature of J0437 can be reached is with $f\lesssim f^*$, as illustrated for $\Delta_n^\infty=0$ in Sect.~\ref{sec:MSP normal neutrons}. In this case, the total reaction rate at $\eta_\ell^\infty\gtrsim\Delta_{\text{thr}}^{\rm in,\infty}=\Delta_n^\infty$ remains small enough for the chemical imbalances to keep growing beyond this value. In the extreme case of $f=0$, we recover the case with no magnetization, in which the imbalances grow to $\eta_\ell^\infty\sim\Delta_{\text{thr}}^{\rm out,\infty}$, which has to be $\sim 1.5\,\mathrm{MeV}$ to reproduce the temperature of J0437, also satisfying the upper limit for J2124.

We conclude that, for MSPs with superconducting protons and superfluid neutrons in their core, if $f\lesssim f^*$, the chemical imbalances equilibrate at $\eta_\ell^\infty\sim\Delta_{\mathrm{thr}}^{\rm out,\infty}$ (given by Eq.~\ref{thresh}), whereas for $f\gtrsim f^*$ they equilibrate at $\eta_\ell^\infty\sim\Delta_{\mathrm{thr}}^{\rm in,\infty}=\Delta_n^\infty$. For some intermediate values $f\sim f^*$, the chemical imbalances will equilibrate at an intermediate value, $\Delta_{\mathrm{thr}}^{\rm in,\infty}\lesssim \eta_\ell^\infty\lesssim\Delta_{\mathrm{thr}}^{\rm out,\infty}$. In all cases, the {quasi-steady} surface temperature is directly linked to the {quasi-steady} chemical imbalances, requiring the latter to be $\eta_\ell^\infty\approx 1.5\,\mathrm{MeV}$ to account for the measured temperature of J0437.

\begin{figure}
\includegraphics[width=1\linewidth]{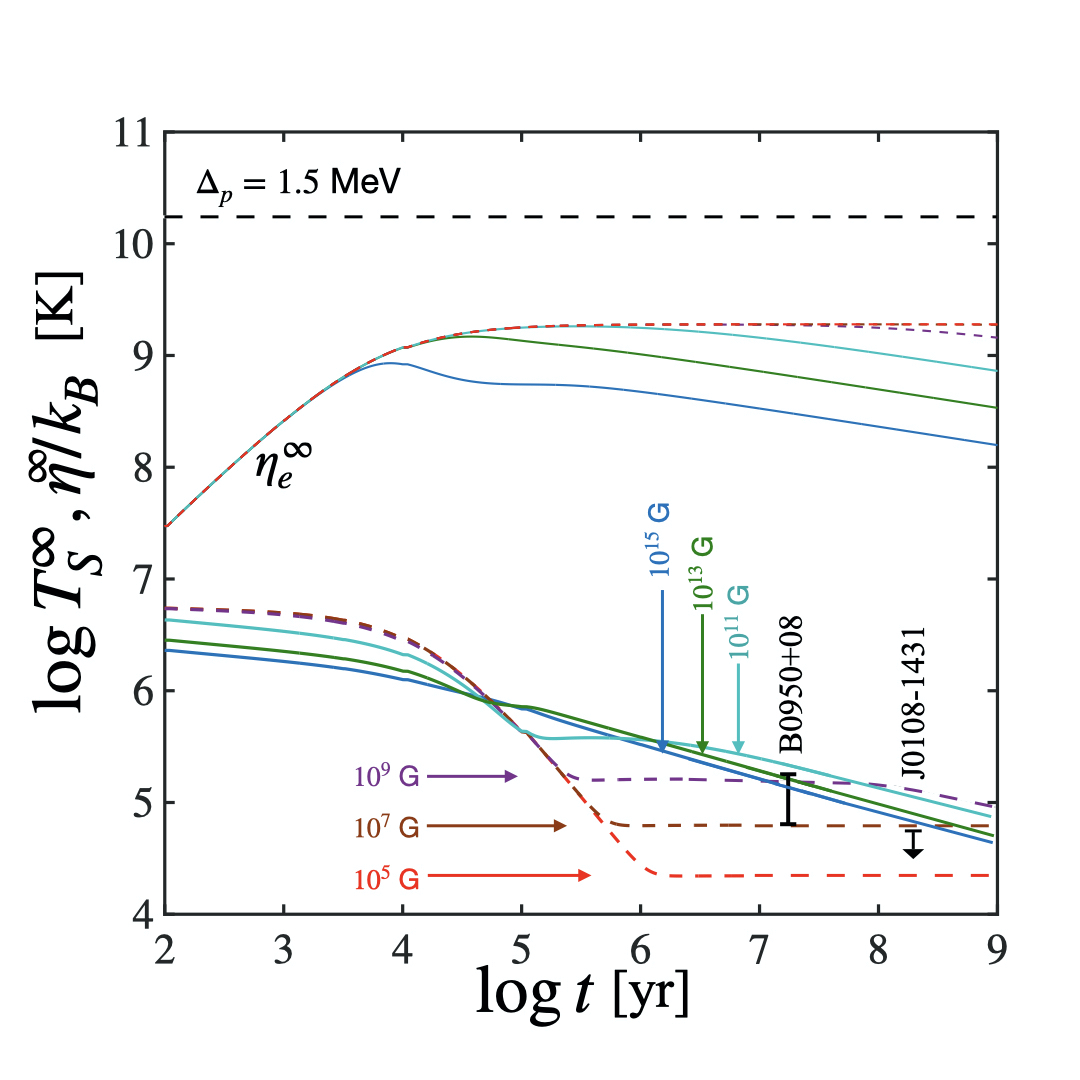}
\caption{{Evolution of the redshifted surface temperature ($T_{\rm s}^\infty$; lower curves) and electron chemical imbalance ($\eta_e^\infty$; upper curves) for rotochemical heating with normal neutrons and superconducting protons with a uniform proton Cooper pairing gap $\Delta_p=1.5$ MeV, assuming only Murca reactions are allowed. The spin-down evolution was calculated with the magnetic dipole moment inferred for PSR~B0950+08, which is almost identical to that of PSR~J0108-1431; the temperature measurement for the former and the upper limit for the latter are shown. The blue, green, cyan, purple, brown, and red curves consider core magnetic flux densities $B_{\rm int}=10^{15}, 10^{13}, 10^{11}, 10^9, 10^7$, and $10^5$ G, respectively, and a critical magnetic field $H_{\rm crit}=10^{15}$ G. (Strong $B_{\rm int}$ are shown with continuous lines, weak $B_{\rm int}$ with segmented lines, in order to emphasize the two different regimes discussed in the text.) For clarity, we do not show the curves for $\eta_\mu^\infty/k_B$, which are very similar to those for $\eta_e^\infty/k_B$.} All curves were computed considering an initial rotation period $P_0=5$~ms, initial core temperature $T_0^\infty=10^{11}$~K, equation of state A18+$\delta v$+UIX$^*$ \citep{Akmal1998}, mass $M=1.44\,M_{\odot}$, and coordinate radius $R=11.45\,\mathrm{km}$. 
}\label{BintCL}
\end{figure}

\begin{figure}
\includegraphics[width=1\linewidth]{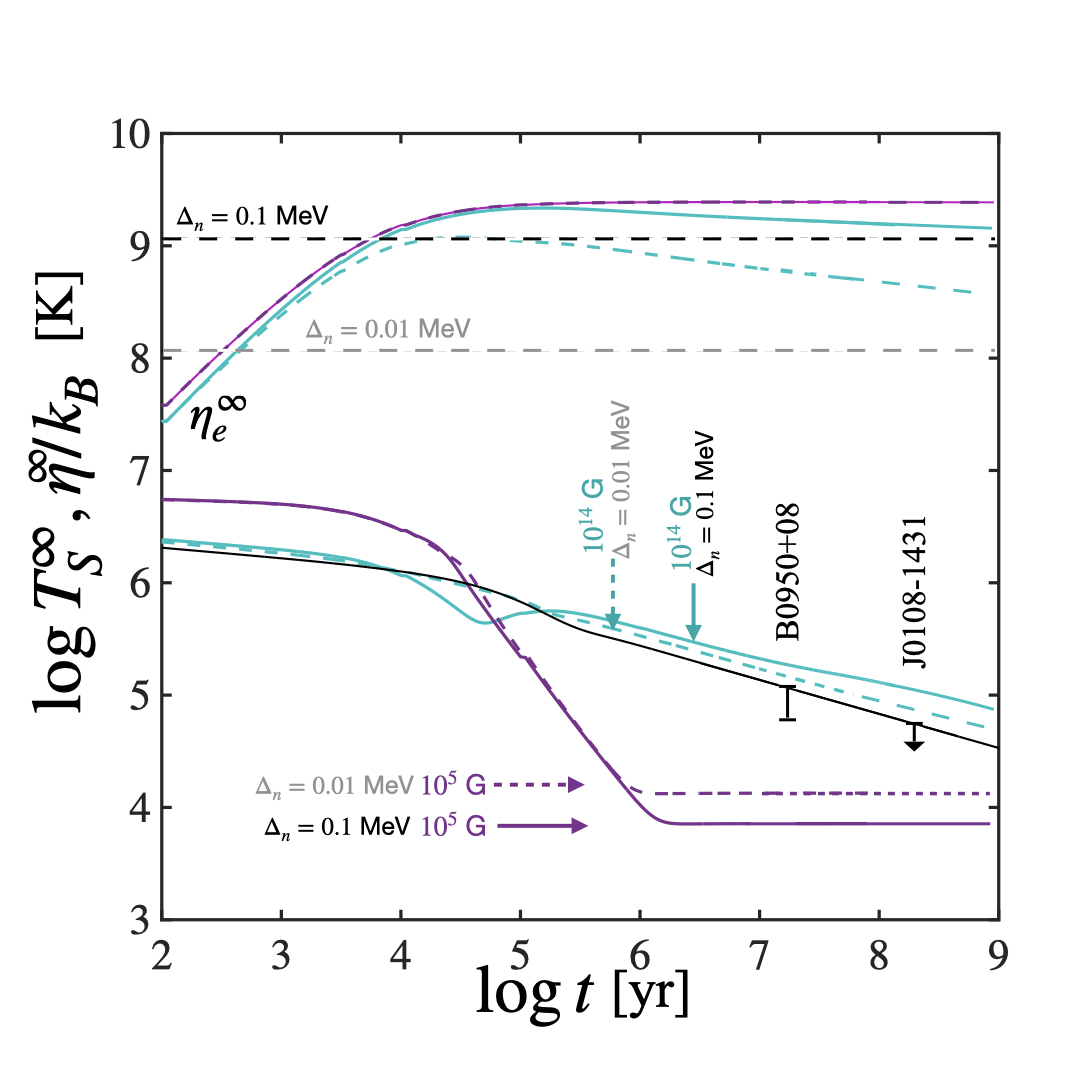}
\caption{{The same as Fig.~\ref{BintCL}, but considering superfluid neutrons and superconducting protons. For reference, the black line corresponds to rotochemical heating with normal protons and normal neutrons. For all other curves, $\Delta_p=1.5$ MeV is used for the superconducting protons, while two different gaps are considered for the superfluid neutrons:} $\Delta_n = 0.01$ MeV (segmented lines) and $\Delta_n=0.1$ MeV (solid lines). The purple lines represent a weak internal magnetic field regime with $B_{\rm int}=10^5$ G, and the green lines represent a strong internal magnetic field regime with $B_{\rm int}=10^{14}$ G, assuming $H_{\rm crit}=10^{15}\,\mathrm{G}$}. 
\label{BintneutronCLP}
\end{figure}

\subsection{Classical pulsars}

{For CPs, the regimes 
in which 
superconducting protons 
coexist with either normal or superfluid neutrons are 
shown in Figs.~\ref{BintCL} and ~\ref{BintneutronCLP}, respectively. 
In both cases, the chemical imbalances initially increase quickly, due to the strong spin-down and unimpeded by Urca reactions, and eventually stop growing once the rotation has substantially slowed down.}


\subsubsection{Superconducting protons and normal neutrons}

{For normal neutrons and superconducting protons with a very small $f$, 
the effect of the reactions on the chemical imbalances always remains negligible and their asymptotic value depends only on the initial rotation rate of the star, $\eta_\ell^{eq}\approx W_{np\ell}\Omega_0^2$, where the constant factor $W_{np\ell}$, defined in \citet{Fer2005}, depends on the NS mass and equation of state. Nevertheless, the few occurring reactions deposit heat at a net rate 
\begin{equation}\label{WeakBintCL}
    L_\gamma^\infty
    \approx\sum_\ell h_\ell\eta_\ell^\infty\Delta\tilde\Gamma_{np\ell}^\infty\approx f\sum_\ell h_\ell A_\ell(W_{np\ell}\Omega_0^2)^{n+1},
\end{equation}
corresponding to} the flat pieces of the curves with $B_{\rm int}=10^5\mathrm{G}$, $10^7\mathrm{G}$, and $10^9\mathrm{G}$ in Fig.~\ref{BintCL}. 

In the opposite regime of large $f$, the initial growth of the imbalances is still fully determined by the spin-down of the NS, but eventually the effect of the reactions becomes stronger, while the spin-down effect becomes weaker, and the late-time evolution corresponds to a decay of the imbalances completely governed by the reactions, at a rate $\dot\eta_\ell^\infty\propto-f(\eta_\ell^\infty)^n$, yielding a luminosity evolution
\begin{equation}\label{StrongBintCL}
    L_\gamma^\infty(t)\propto f^{-2/(n-1)}t^{-(n+1)/(n-1)}.
\end{equation}
{Thus, as also seen in Fig.~\ref{BintCL}, the luminosity at a given time $t\gtrsim 10^6\,\mathrm{yr}$ increases $\propto f$ for small values of $f$ (Eq.~\ref{WeakBintCL}) 
and decreases as $f^{-2/(n-1)}$ for large values (Eq.~\ref{StrongBintCL}), so the maximum luminosity is obtained for $f$ around the transition between these regimes. We note that the evolution is independent of $\Delta_p$, as long as the latter is large enough to always have $\eta_\ell\ll\Delta_p$.}

It can be seen that cases with large $f$ match the observation of B0950 and are almost consistent with the upper limit inferred for J0108. Cases with small $f$ can match both observations for different combinations of $f$ and $P_0$.

\subsubsection{Superconducting protons and superfluid neutrons}

{When the superconducting protons are joined by superfluid neutrons with a moderate gap, $\Delta_n^\infty\lesssim 0.1\,\mathrm{MeV}$, as illustrated by Fig.~\ref{BintneutronCLP}, the regimes of small and large $f$ are slightly modified. In the former, the evolution of the chemical imbalances is again unaffected by the reactions, yielding the same asymptotic value as for normal neutrons, $\eta_\ell^\infty\approx W_{np\ell}\Omega_0^2$. However, now the few reactions occurring in the magnetized regions are further reduced by the neutron Cooper pairing, making the asymptotic temperature a decreasing function of $\Delta_n^\infty$, as seen by comparing the curves with the same $f=10^{-10}$ and two different values of $\Delta_n^\infty=0.1$ MeV and $\Delta_n^\infty=0.01$ MeV. This case can easily fit the observations for appropriate combinations of $f$, $\Omega_0$, and $\Delta_n^\infty$.

If, on the other hand, $f$ is large enough, the reactions stop the growth of the chemical potentials soon after the latter cross the threshold $\eta_\ell^\infty\sim\Delta_{\mathrm{thr}}^{\rm in,\infty}=\Delta_n^\infty$, therefore the chemical imbalances and the surface temperature become increasing functions of $\Delta_n^\infty$, as shown for the set of lines with $f=10^{-1}$.} In this limit, the observations cannot be fitted if taken at face value, although they are close to the predictions for small $\Delta_n^\infty$.

\begin{figure}
\includegraphics[width=1\linewidth]{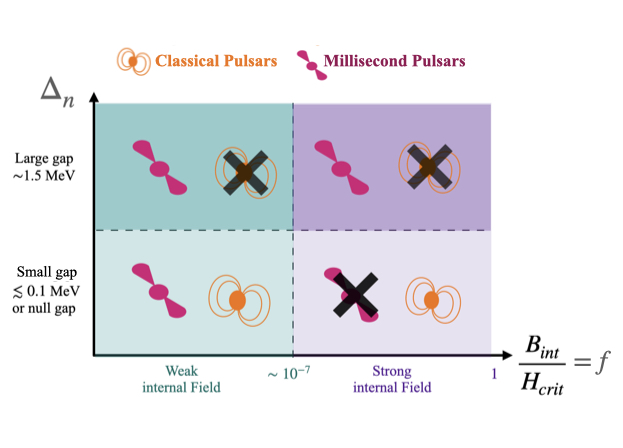}
\caption{A schematic diagram illustrating the impact of rotochemical heating in the presence of superconducting protons with an energy gap of $\Delta_p=1.5$ MeV, for different magnitudes of the internal magnetic field $B_{\rm int}$ and neutron energy gap $\Delta_n$. Two neutron star icons are depicted: the orange one represents CPs, and the purple one represents MSPs. These icons denote regimes of $B_{\rm int}$ and $\Delta_n$ under which this process can explain the observed temperatures, including their upper limits, as inferred from observations.}
\label{esquema}
\end{figure}

\section{Summary and discussion}
\label{sec:summary}

Our results are summarized in Fig.~\ref{esquema}, which provides a comprehensive overview of the outcomes resulting from various combinations of the internal magnetic field and the neutron energy gap, while maintaining a constant energy gap $\Delta_p^\infty=1.5$ MeV for the superconducting protons. 

For a large $\Delta_n^\infty\sim 1.5\,\mathrm{MeV}$, CPs require unrealistically short initial periods to reach the threshold in the magnetized regions, therefore they do not trigger reactions regardless of the internal magnetic field strength (see R25). Moreover, simulating CPs with an unrealistically short initial period $P_0 \sim 1$ ms, we observed that in this case the predicted temperatures overestimated the observed ones by a factor $\sim 10$ (not shown). 
On the other hand, our model with $\Delta_n^\infty\sim 1.5\,\mathrm{MeV}$ does explain the observations of MSPs regardless of the internal magnetic field. 

For a small neutron energy gap, rotochemical heating in CPs can explain the observations. The beta reactions are always triggered in the magnetized regions, and, by tuning the initial period, the internal magnetic field, and the neutron energy gap, it is always possible to fit the inferred temperatures.

On the other hand, MSPs with a small $\Delta_n^\infty$ require a very weak internal field. This is because a large $f\approx B_{\rm int}/H_{\rm crit}$ allows Urca reactions to consume chemical energy inside magnetized regions since relatively early stages, reaching a colder {quasi-steady-state} temperature. A weak magnetic field prevents the premature consumption of energy inside magnetized regions, making MSPs able to reach $\Delta_{\rm thr}^{\rm out,\infty}$ and trigger reactions in the whole core (if $\Delta_n^\infty \ll \Delta_p^\infty\sim 1.5$ MeV).

Thus, considering that CPs and MSPs should have roughly the same $\Delta_p^\infty$ and $\Delta_n^\infty$, but possibly different $f$ and almost certainly different $P_0$ (even within each class), the only viable option is a large $\Delta_p^\infty\approx 1.5\,\mathrm{MeV}$ and a relatively small or vanishing $\Delta_n^\infty\lesssim 0.1\,\mathrm{MeV}$. For MSPs, this must be accompanied by a very small $f$, whereas for CPs both small and large values of $f$ are acceptable.

The required value of $\Delta_p^\infty\approx 1.5\,\mathrm{MeV}$ is somewhat larger than expected from theoretical models (see, e.~g., \citealt{Potekhin2026}), although large values have been proposed to explain the fast cooling observed in the NS in the Cassiopeia A supernova remnant \citep{Page2011,Shternin2011}. The small value of $f$ in MSPs would imply that, in these objects, the magnetic field is almost completely expelled from the NS core.

However, we must note that our conclusions depend on our assumption that there are no heating mechanisms besides rotochemical heating. If vortex creep is also allowed, one finds that the conditions for rotochemical heating can be relaxed. In particular, vortex creep could explain the high temperature of B0950, while rotochemical heating with a large neutron gap $\Delta_n^\infty\approx 1.5\,\mathrm{MeV}$ can explain that of J0437 (R25), regardless of the value of $f$. 

\begin{acknowledgements}
This work was supported by ANID-FONDECYT grant 1201582.
\end{acknowledgements}

\bibliographystyle{aa}
\bibliography{biblio}

@ARTICLE{Abramkin2021,
       author = {{Abramkin}, Vadim and {Shibanov}, Yuriy and {Mignani}, Roberto P. and {Pavlov}, George G.},
        title = "{Hubble Space Telescope Observations of the Old Pulsar PSR J0108-1431}",
      journal = {\apj},
         year = 2021,
        month = apr,
       volume = {911},
       number = {1},
          eid = {1},
        pages = {1},
          doi = {10.3847/1538-4357/abe704},
archivePrefix = {arXiv},
       eprint = {2103.00332},
 primaryClass = {astro-ph.HE},
       adsurl = {https://ui.adsabs.harvard.edu/abs/2021ApJ...911....1A}
}

@ARTICLE{Abramkin2022,
       author = {{Abramkin}, Vadim and {Pavlov}, George G. and {Shibanov}, Yuriy and {Kargaltsev}, Oleg},
        title = "{Thermal and Nonthermal Emission in the Optical-UV Spectrum of PSR B0950+08}",
      journal = {\apj},
         year = 2022,
        month = jan,
       volume = {924},
       number = {2},
          eid = {128},
        pages = {128},
          doi = {10.3847/1538-4357/ac3a6f},
archivePrefix = {arXiv},
       eprint = {2111.08801},
 primaryClass = {astro-ph.HE},
       adsurl = {https://ui.adsabs.harvard.edu/abs/2022ApJ...924..128A}
}

@ARTICLE{Akmal1998,
       author = {{Akmal}, A. and {Pandharipande}, V.~R. and {Ravenhall}, D.~G.},
        title = "{Equation of state of nucleon matter and neutron star structure}",
      journal = {\prc},
         year = 1998,
        month = sep,
       volume = {58},
       number = {3},
        pages = {1804-1828},
          doi = {10.1103/PhysRevC.58.1804},
archivePrefix = {arXiv},
       eprint = {nucl-th/9804027},
 primaryClass = {nucl-th},
       adsurl = {https://ui.adsabs.harvard.edu/abs/1998PhRvC..58.1804A}
}

@ARTICLE{Alpar1984,
   author = {{Alpar}, M.~A. and {Pines}, D. and {Anderson}, P.~W. and {Shaham}, J.
	},
    title = "{Vortex creep and the internal temperature of neutron stars. I - General theory}",
  journal = {\apj},
     year = 1984,
    month = jan,
   volume = 276,
    pages = {325-334},
      doi = {10.1086/161616},
   adsurl = {https://ui.adsabs.harvard.edu/abs/1984ApJ...276..325A}
}

@ARTICLE{Baym71,
       author = {{Baym}, G. and {Pines}, D.},
        title = "{Neutron starquakes and pulsar speedup.}",
      journal = {Annals of Physics},
         year = 1971,
        month = jan,
       volume = {66},
        pages = {816-835},
          doi = {10.1016/0003-4916(71)90084-4},
       adsurl = {https://ui.adsabs.harvard.edu/abs/1971AnPhy..66..816B}
}

@ARTICLE{Be2018,
       author = {{Beloin}, Spencer and {Han}, Sophia and {Steiner}, Andrew W. and {Page}, Dany},
        title = "{Constraining superfluidity in dense matter from the cooling of isolated neutron stars}",
      journal = {\prc},
         year = 2018,
        month = jan,
       volume = {97},
       number = {1},
          eid = {015804},
        pages = {015804},
          doi = {10.1103/PhysRevC.97.015804},
       adsurl = {https://ui.adsabs.harvard.edu/abs/2018PhRvC..97a5804B}
}

@ARTICLE{Camilo1994,
       author = {{Camilo}, F. and {Thorsett}, S.~E. and {Kulkarni}, S.~R.},
        title = "{The Magnetic Fields, Ages, and Original Spin Periods of Millisecond Pulsars}",
      journal = {\apjl},
         year = "1994",
        month = "Jan",
       volume = {421},
        pages = {L15},
          doi = {10.1086/187176},
       adsurl = {https://ui.adsabs.harvard.edu/abs/1994ApJ...421L..15C}
}

@ARTICLE{Cheng1992,
       author = {{Cheng}, K.~S. and {Chau}, W.~Y. and {Zhang}, J.~L. and {Chau}, H.~F.},
        title = "{Effects of Evolving Rotating Equilibrium Configurations on the Cooling and Spin-down of Pulsars}",
      journal = {\apj},
         year = 1992,
        month = sep,
       volume = {396},
        pages = {135},
          doi = {10.1086/171704},
       adsurl = {https://ui.adsabs.harvard.edu/abs/1992ApJ...396..135C}
}

@ARTICLE{deLava,
       author = {{de Lavallaz}, Arnaud and {Fairbairn}, Malcolm},
        title = "{Neutron stars as dark matter probes}",
      journal = {\prd},
         year = 2010,
        month = jun,
       volume = {81},
       number = {12},
          eid = {123521},
        pages = {123521},
          doi = {10.1103/PhysRevD.81.123521},
archivePrefix = {arXiv},
       eprint = {1004.0629},
 primaryClass = {astro-ph.GA},
       adsurl = {https://ui.adsabs.harvard.edu/abs/2010PhRvD..81l3521D}
}

@ARTICLE{Durant2012,
   author = {{Durant}, M. and {Kargaltsev}, O. and {Pavlov}, G.~G. and {Kowalski}, P.~M. and 
	{Posselt}, B. and {van Kerkwijk}, M.~H. and {Kaplan}, D.~L.},
  journal = {\apj},
archivePrefix = "arXiv",
   eprint = {1111.2346},
 primaryClass = "astro-ph.HE",
     year = 2012,
    month = feb,
   volume = 746,
      eid = {6},
    pages = {6},
      doi = {10.1088/0004-637X/746/1/6},
   adsurl = {https://ui.adsabs.harvard.edu/abs/2012ApJ...746....6D}
}

@ARTICLE{Fer2005,
   author = {{Fern{\'a}ndez}, R. and {Reisenegger}, A.},
  journal = {\apj},
   eprint = {astro-ph/0502116},
     year = 2005,
    month = may,
   volume = 625,
    pages = {291-306},
      doi = {10.1086/429551},
   adsurl = {https://ui.adsabs.harvard.edu/abs/2005ApJ...625..291F}
}

@ARTICLE{Gold92,
       author = {{Goldreich}, Peter and {Reisenegger}, Andreas},
        title = "{Magnetic Field Decay in Isolated Neutron Stars}",
      journal = {\apj},
         year = 1992,
        month = aug,
       volume = {395},
        pages = {250},
          doi = {10.1086/171646},
       adsurl = {https://ui.adsabs.harvard.edu/abs/1992ApJ...395..250G}
}

@ARTICLE{GonRei2010,
   author = {{Gonzalez}, D. and {Reisenegger}, A.},
  journal = {\aap},
archivePrefix = "arXiv",
   eprint = {1005.5699},
 primaryClass = "astro-ph.HE",
     year = 2010,
    month = nov,
   volume = 522,
      eid = {A16},
    pages = {A16},
      doi = {10.1051/0004-6361/201015084},
   adsurl = {https://ui.adsabs.harvard.edu/abs/2010A%26A...522A..16G}
}

@ARTICLE{GonJ2015,
   author = {{Gonz{\'a}lez-Jim{\'e}nez}, N. and {Petrovich}, C. and {Reisenegger}, A.
	},
  journal = {\mnras},
archivePrefix = "arXiv",
   eprint = {1411.6500},
 primaryClass = "astro-ph.SR",
     year = 2015,
    month = mar,
   volume = 447,
    pages = {2073-2084},
      doi = {10.1093/mnras/stu2558},
   adsurl = {https://ui.adsabs.harvard.edu/abs/2015MNRAS.447.2073G}
}

@ARTICLE{Gon2019,
       author = {{Gonz{\'a}lez-Caniulef}, Denis and {Guillot}, Sebastien and
         {Reisenegger}, Andreas},
        title = "{Neutron star radius measurement from the ultraviolet and soft X-ray thermal emission of PSR J0437-4715}",
      journal = {\mnras},
         year = 2019,
        month = dec,
       volume = {490},
       number = {4},
        pages = {5848-5859},
          doi = {10.1093/mnras/stz2941},
archivePrefix = {arXiv},
       eprint = {1904.12114},
 primaryClass = {astro-ph.HE},
       adsurl = {https://ui.adsabs.harvard.edu/abs/2019MNRAS.490.5848G}
}

@ARTICLE{Gud83,
   author = {{Gudmundsson}, E.~H. and {Pethick}, C.~J. and {Epstein}, R.~I.
	},
    title = "{Structure of neutron star envelopes}",
  journal = {\apj},
     year = 1983,
    month = sep,
   volume = 272,
    pages = {286-300},
      doi = {10.1086/161292},
   adsurl = {https://ui.adsabs.harvard.edu/abs/1983ApJ...272..286G}
}

@ARTICLE{Guillot2019,
   author = {{Guillot}, S. and {Pavlov}, G.~G. and {Reyes}, C. and {Reisenegger}, A. and 
	{Rodriguez}, L.~E. and {Rangelov}, B. and {Kargaltsev}, O.},
  journal = {\apj},
archivePrefix = "arXiv",
   eprint = {1901.07998},
 primaryClass = "astro-ph.HE",
     year = 2019,
    month = apr,
   volume = 874,
      eid = {175},
    pages = {175},
      doi = {10.3847/1538-4357/ab0f38},
   adsurl = {https://ui.adsabs.harvard.edu/abs/2019ApJ...874..175G}
}

@ARTICLE{Kantor2021,
       author = {{Kantor}, E.~M. and {Gusakov}, M.~E.},
        title = "{Long-lasting accretion-powered chemical heating of millisecond pulsars}",
      journal = {\mnras},
         year = 2021,
        month = dec,
       volume = {508},
       number = {4},
        pages = {6118-6127},
          doi = {10.1093/mnras/stab2922},
archivePrefix = {arXiv},
       eprint = {2110.02881},
 primaryClass = {astro-ph.HE},
       adsurl = {https://ui.adsabs.harvard.edu/abs/2021MNRAS.508.6118K}
}

@ARTICLE{Kar2004,
   author = {{Kargaltsev}, O. and {Pavlov}, G.~G. and {Romani}, R.~W.},
    title = "{Ultraviolet Emission from the Millisecond Pulsar J0437-4715}",
  journal = {\apj},
   eprint = {astro-ph/0310854},
     year = 2004,
    month = feb,
   volume = 602,
    pages = {327-335},
      doi = {10.1086/380993},
   adsurl = {https://ui.adsabs.harvard.edu/abs/2004ApJ...602..327K}
}

@ARTICLE{Kou10,
       author = {{Kouvaris}, Chris and {Tinyakov}, Peter},
        title = "{Can neutron stars constrain dark matter?}",
      journal = {\prd},
         year = 2010,
        month = sep,
       volume = {82},
       number = {6},
          eid = {063531},
        pages = {063531},
          doi = {10.1103/PhysRevD.82.063531},
archivePrefix = {arXiv},
       eprint = {1004.0586},
 primaryClass = {astro-ph.GA},
       adsurl = {https://ui.adsabs.harvard.edu/abs/2010PhRvD..82f3531K}
}

@ARTICLE{Kopp2023,
       author = {{K{\"o}pp}, F. and {Horvath}, J.~E. and {Hadjimichef}, D. and {Vasconcellos}, C.~A.~Z. and {Hess}, P.~O.},
        title = "{Internal heating mechanisms in neutron stars}",
      journal = {International Journal of Modern Physics D},
         year = 2023,
        month = jan,
       volume = {32},
       number = {7},
          eid = {2350046-51},
        pages = {2350046-51},
          doi = {10.1142/S0218271823500463},
archivePrefix = {arXiv},
       eprint = {2208.07770},
 primaryClass = {astro-ph.HE},
       adsurl = {https://ui.adsabs.harvard.edu/abs/2023IJMPD..3250046K}
}

@ARTICLE{Page2011,
       author = {{Page}, Dany and {Prakash}, Madappa and {Lattimer}, James M. and {Steiner}, Andrew W.},
        title = "{Rapid Cooling of the Neutron Star in Cassiopeia A Triggered by Neutron Superfluidity in Dense Matter}",
      journal = {\prl},
         year = 2011,
        month = feb,
       volume = {106},
       number = {8},
          eid = {081101},
        pages = {081101},
          doi = {10.1103/PhysRevLett.106.081101},
archivePrefix = {arXiv},
       eprint = {1011.6142},
 primaryClass = {astro-ph.HE},
       adsurl = {https://ui.adsabs.harvard.edu/abs/2011PhRvL.106h1101P}
}

@ARTICLE{Pavlov2017,
   author = {{Pavlov}, G.~G. and {Rangelov}, B. and {Kargaltsev}, O. and 
	{Reisenegger}, A. and {Guillot}, S. and {Reyes}, C.},
    title = "{Old but Still Warm: Far-UV Detection of PSR B0950+08}",
  journal = {\apj},
archivePrefix = "arXiv",
   eprint = {1710.06448},
 primaryClass = "astro-ph.HE",
     year = 2017,
    month = nov,
   volume = 850,
      eid = {79},
    pages = {79},
      doi = {10.3847/1538-4357/aa947c},
   adsurl = {https://ui.adsabs.harvard.edu/abs/2017ApJ...850...79P}
}

@ARTICLE{Petro2010,
   author = {{Petrovich}, C. and {Reisenegger}, A.},
    title = "{Rotochemical heating in millisecond pulsars: modified Urca reactions with uniform Cooper pairing gaps}",
  journal = {\aap},
archivePrefix = "arXiv",
   eprint = {0912.2564},
 primaryClass = "astro-ph.HE",
     year = 2010,
    month = oct,
   volume = 521,
      eid = {A77},
    pages = {A77},
      doi = {10.1051/0004-6361/200913861},
   adsurl = {https://ui.adsabs.harvard.edu/abs/2010A%26A...521A..77P}
}

@ARTICLE{Petro2011,
   author = {{Petrovich}, C. and {Reisenegger}, A.},
    title = "{Long-period thermal oscillations in superfluid millisecond pulsars}",
  journal = {\aap},
archivePrefix = "arXiv",
   eprint = {1008.3013},
 primaryClass = "astro-ph.HE",
     year = 2011,
    month = apr,
   volume = 528,
      eid = {A66},
    pages = {A66},
      doi = {10.1051/0004-6361/201015603},
   adsurl = {https://ui.adsabs.harvard.edu/abs/2011A%26A...528A..66P}
}

@ARTICLE{Pons2025,
       author = {{Pons}, Jos{\'e} A. and {Dehman}, Clara and {Vigan{\`o}}, Daniele},
        title = "{Magnetic, thermal and rotational evolution of isolated neutron stars}",
      journal = {arXiv e-prints},
         year = 2025,
        month = sep,
          eid = {arXiv:2509.06699},
        pages = {arXiv:2509.06699},
          doi = {10.48550/arXiv.2509.06699},
archivePrefix = {arXiv},
       eprint = {2509.06699},
 primaryClass = {astro-ph.HE},
       adsurl = {https://ui.adsabs.harvard.edu/abs/2025arXiv250906699P}
}

@ARTICLE{Potekhin1997,
       author = {{Potekhin}, A.~Y. and {Chabrier}, G. and {Yakovlev}, D.~G.},
        title = "{Internal temperatures and cooling of neutron stars with accreted envelopes.}",
      journal = {\aap},
         year = 1997,
        month = jul,
       volume = {323},
        pages = {415-428},
archivePrefix = {arXiv},
       eprint = {astro-ph/9706148},
 primaryClass = {astro-ph},
       adsurl = {https://ui.adsabs.harvard.edu/abs/1997A&A...323..415P}
}

@ARTICLE{Potekhin2026,
       author = {{Potekhin}, A.~Y. and {Yakovlev}, D.~G.},
        title = "{Urca cooling of the neutron star in the Cassiopeia A supernova remnant}",
      journal = {Journal of High Energy Astrophysics},
         year = 2026,
        month = jan,
       volume = {49},
          eid = {100441},
        pages = {100441},
          doi = {10.1016/j.jheap.2025.100441},
       adsurl = {https://ui.adsabs.harvard.edu/abs/2026JHEAp..4900441P}
}

@ARTICLE{Range2017,
   author = {{Rangelov}, B. and {Pavlov}, G.~G. and {Kargaltsev}, O. and 
	{Reisenegger}, A. and {Guillot}, S. and {van Kerkwijk}, M.~H. and 
	{Reyes}, C.},
    title = "{Hubble Space Telescope Detection of the Millisecond Pulsar J2124-3358 and its Far-ultraviolet Bow Shock Nebula}",
  journal = {\apj},
archivePrefix = "arXiv",
   eprint = {1701.00002},
 primaryClass = "astro-ph.HE",
     year = 2017,
    month = feb,
   volume = 835,
      eid = {264},
    pages = {264},
      doi = {10.3847/1538-4357/835/2/264},
   adsurl = {https://ui.adsabs.harvard.edu/abs/2017ApJ...835..264R}
}

@ARTICLE{Reis1995,
   author = {{Reisenegger}, A.},
   title = "{Deviations from chemical equilibrium due to spin-down as an internal heat source in neutron stars}",
  journal = {\apj},
   eprint = {astro-ph/9410035},
     year = 1995,
    month = apr,
   volume = 442,
    pages = {749-757},
      doi = {10.1086/175480},
   adsurl = {https://ui.adsabs.harvard.edu/abs/1995ApJ...442..749R}
}

@ARTICLE{Reis1997,
       author = {{Reisenegger}, Andreas},
        title = "{Constraining Dense Matter Superfluidity through Thermal Emission from Millisecond Pulsars}",
      journal = {\apj},
         year = "1997",
        month = "Aug",
       volume = {485},
       number = {1},
        pages = {313-318},
          doi = {10.1086/304417},
archivePrefix = {arXiv},
       eprint = {astro-ph/9612179},
 primaryClass = {astro-ph},
       adsurl = {https://ui.adsabs.harvard.edu/abs/1997ApJ...485..313R}
}

@ARTICLE{Reisenegger2006,
       author = {{Reisenegger}, Andreas and {Jofr{\'e}}, Paula and {Fern{\'a}ndez}, Rodrigo and {Kantor}, Elena},
        title = "{Rotochemical Heating of Neutron Stars: Rigorous Formalism with Electrostatic Potential Perturbations}",
      journal = {\apj},
         year = 2006,
        month = dec,
       volume = {653},
       number = {1},
        pages = {568-572},
          doi = {10.1086/506601},
archivePrefix = {arXiv},
       eprint = {astro-ph/0606322},
 primaryClass = {astro-ph},
       adsurl = {https://ui.adsabs.harvard.edu/abs/2006ApJ...653..568R}
}

@ARTICLE{Rodriguez2025,
       author = {{Rodr{\'\i}guez}, Luis E. and {Reisenegger}, Andreas and {Gonz{\'a}lez-Caniulef}, Denis and {Petrovich}, Crist{\'o}bal and {Pavlov}, George and {Guillot}, S{\'e}bastien and {Kargaltsev}, Oleg and {Rangelov}, Blagoy},
        title = "{Neutron star heating vs. HST observations}",
      journal = {arXiv e-prints},
         year = 2025,
        month = nov,
          eid = {arXiv:2511.16507},
        pages = {arXiv:2511.16507},
          doi = {10.48550/arXiv.2511.16507},
archivePrefix = {arXiv},
       eprint = {2511.16507},
 primaryClass = {astro-ph.HE},
       adsurl = {https://ui.adsabs.harvard.edu/abs/2025arXiv251116507R}
}

@ARTICLE{Schaab1998,
       author = {{Schaab}, Ch. and {Weber}, F. and {Weigel}, M.~K.},
        title = "{Neutron superfluidity in strongly magnetic interiors of neutron stars and its effect on thermal evolution}",
      journal = {\aap},
         year = 1998,
        month = jul,
       volume = {335},
        pages = {596-604},
archivePrefix = {arXiv},
       eprint = {astro-ph/9804189},
 primaryClass = {astro-ph},
       adsurl = {https://ui.adsabs.harvard.edu/abs/1998A&A...335..596S}
}

@ARTICLE{Sedrakian2005,
       author = {{Sedrakian}, Armen},
        title = "{Type-I superconductivity and neutron star precession}",
      journal = {\prd},
         year = 2005,
        month = apr,
       volume = {71},
       number = {8},
          eid = {083003},
        pages = {083003},
          doi = {10.1103/PhysRevD.71.083003},
       adsurl = {https://ui.adsabs.harvard.edu/abs/2005PhRvD..71h3003S}
}

@ARTICLE{Shklovskii1970,
       author = {{Shklovskii}, I.~S.},
        title = "{Possible Causes of the Secular Increase in Pulsar Periods.}",
      journal = {\sovast},
         year = "1970",
        month = "Feb",
       volume = {13},
        pages = {562},
       adsurl = {https://ui.adsabs.harvard.edu/abs/1970SvA....13..562S}
}

@ARTICLE{Shternin2011,
       author = {{Shternin}, Peter S. and {Yakovlev}, Dmitry G. and {Heinke}, Craig O. and {Ho}, Wynn C.~G. and {Patnaude}, Daniel J.},
        title = "{Cooling neutron star in the Cassiopeia A supernova remnant: evidence for superfluidity in the core}",
      journal = {\mnras},
         year = 2011,
        month = mar,
       volume = {412},
       number = {1},
        pages = {L108-L112},
          doi = {10.1111/j.1745-3933.2011.01015.x},
archivePrefix = {arXiv},
       eprint = {1012.0045},
 primaryClass = {astro-ph.SR},
       adsurl = {https://ui.adsabs.harvard.edu/abs/2011MNRAS.412L.108S}
}

@ARTICLE{Thom96,
       author = {{Thompson}, Christopher and {Duncan}, Robert C.},
        title = "{The Soft Gamma Repeaters as Very Strongly Magnetized Neutron Stars. II. Quiescent Neutrino, X-Ray, and Alfven Wave Emission}",
      journal = {\apj},
         year = 1996,
        month = dec,
       volume = {473},
        pages = {322},
          doi = {10.1086/178147},
       adsurl = {https://ui.adsabs.harvard.edu/abs/1996ApJ...473..322T}
}

@ARTICLE{Thorne1977,
       author = {{Thorne}, K.~S.},
        title = "{The relativistic equations of stellar structure and evolution.}",
      journal = {\apj},
         year = "1977",
        month = "Mar",
       volume = {212},
        pages = {825-831},
          doi = {10.1086/155108},
       adsurl = {https://ui.adsabs.harvard.edu/abs/1977ApJ...212..825T}
}

@BOOK{Tilley1990,
   author = {{Tilley}, D.~R. and {Tilley}, J.},
    title = "{Superfluidity and Superconductivity}",
     year = 1990,
     publisher = {Bristol (IOP)},
     edition = {3rd}
}

@ARTICLE{Yako2004,
   author = {{Yakovlev}, D.~G. and {Pethick}, C.~J.},
    title = "{Neutron Star Cooling}",
  journal = {\araa},
   eprint = {astro-ph/0402143},
     year = 2004,
    month = sep,
   volume = 42,
    pages = {169-210},
      doi = {10.1146/annurev.astro.42.053102.134013},
   adsurl = {https://ui.adsabs.harvard.edu/abs/2004ARA%26A..42..169Y}
}

@ARTICLE{Yanagi2020,
       author = {{Yanagi}, Keisuke and {Nagata}, Natsumi and {Hamaguchi}, Koichi},
        title = "{Cooling theory faced with old warm neutron stars: role of non-equilibrium processes with proton and neutron gaps}",
      journal = {\mnras},
         year = 2020,
        month = mar,
       volume = {492},
       number = {4},
        pages = {5508-5523},
          doi = {10.1093/mnras/staa076},
archivePrefix = {arXiv},
       eprint = {1904.04667},
 primaryClass = {astro-ph.HE},
       adsurl = {https://ui.adsabs.harvard.edu/abs/2020MNRAS.492.5508Y}
}


\end{document}